\documentclass[twocolumn]{aastex631}
\usepackage{xspace} 

\def\casa {\texttt{CASA}\xspace}
\def\ts     {\thinspace}
\def\kms    {\ts km\ts s$^{-1}$}

\def\msol   {$M_{\odot}$}
\def\lsol   {$L_{\odot}$}

\def\oh     {{\rm OH}($^2\Pi_{1/2}\,J$=3/2$\to$1/2)}
\def\cii    {[C{\scriptsize II}]($^2P_{3/2}$$\to$$^2P_{1/2}$)}
\def\oi     {[O{\scriptsize I}]($^3P_{0}$$\to$$^3P_{1}$)}
\def\aco    {{\rm CO}($J$=1$\to$0)}
\def\bco    {{\rm CO}($J$=2$\to$1)}
\def\cco    {{\rm CO}($J$=3$\to$2)}
\def\dco    {{\rm CO}($J$=4$\to$3)}
\def\eco    {{\rm CO}($J$=5$\to$4)}
\def\fco    {{\rm CO}($J$=6$\to$5)}
\def\gco    {{\rm CO}($J$=7$\to$6)}

\def\ico    {{\rm CO}($J$=9$\to$8)}

\def\nco    {{\rm CO}($J$=14$\to$13)}
\def\pco    {{\rm CO}($J$=16$\to$15)}
\def\rco    {{\rm CO}($J$=18$\to$17)}

\received{Jan 10, 2024}
\revised{December 23, 2024}
\accepted{December 30, 2024}

\submitjournal{ApJ}

\shorttitle{Cold Starbursts?}
\shortauthors{Riechers}

\graphicspath{{./}{figures/}}

\begin{document}

\title{Do Unusually Cold Starburst Galaxies Exist? A Case Study.}

\correspondingauthor{Dominik A.\ Riechers}
\email{riechers@ph1.uni-koeln.de}

\author[0000-0001-9585-1462]{Dominik A.\ Riechers}
\affiliation{Institut f\"ur Astrophysik, Universit\"at zu K\"oln, Z\"ulpicher Stra{\ss}e 77, D-50937 K\"oln, Germany}

\begin{abstract}

We report observations of \ico\ and OH$^+$($N$=1$\to$0) toward the
four millimeter-selected lensed starburst galaxies SPT\,2354--58,
0150--59, 0314--44, and 0452--50, using the Atacama Large
Millimeter/submillimeter Array Compact Array (ALMA/ACA), as part of a
larger study of OH$^+$ in the early universe. In this work, we use
these observations for the main purpose of spectroscopic redshift
measurements. For all sources except SPT\,0452--50, we confirm the
previously reported most likely redshifts, and we find typical CO and
OH$^+$ properties for massive starbursts. For SPT\,0452--50, we rule
out the previously reported value of $z$=2.0105, measuring a firm
redshift of $z$=5.0160 based on [OI], [CII], H$_2$O, and CO emission
instead when adding in ancillary ALMA data. Previously, SPT\,0452--50
was considered an outlier in relations between dust temperature,
far-infrared luminosity and redshift, which may have hinted at an
unusually cold starburst with a dust temperature of only $T_{\rm
  dust}$=(21$\pm$2)\,K. Instead, our new measurements suggest it to be
among highly luminous massive dusty starbursts at $z$$>$5, with rather
typical properties within that population. We find a revised dust
temperature of $T_{\rm dust}$=(76.2$\pm$2.5)\,K, and an updated
lensing-corrected far-infrared luminosity (42.5--122.5\,$\mu$m) of
(2.35$^{+0.09}_{-0.08}$)$\times$10$^{13}$\,\lsol\ --
i.e., about an order of magnitude higher than previously reported.  We
thus do not find evidence for the existence of unusually cold
starburst galaxies in the early universe that were missed by previous
selection techniques.

\end{abstract}



\section{Introduction} \label{sec:intro}

Massive dusty starburst galaxies, also commonly labeled ``dusty
star-forming galaxies'' (DSFGs), are an important ingredient to our
understanding of massive galaxy evolution across cosmic history,
because they represent such systems in their most active phases of
growth (see reviews by, e.g., \citealt{blain02,hdc20}). While dust has
now been detected in relatively ``normal'' galaxies back to $z$$>$8
(e.g., \citealt{tamura19}), the space density of DSFGs appears to
drastically decrease toward $z$$>$5, with just over 20
spectroscopically confirmed systems at $z$=5--6 extracted from surveys
extending over thousands of square degrees on the sky, four at
$z$=6--7, and none at earlier epochs (\citealt{riechers10a,riechers13b,riechers14b,riechers17,riechers20a,riechers21a,combes12,walter12b,weiss13,strandet16,strandet17,fudamoto17,zavala18,pavesi18a,jin19,reuter20,ikarashi22,cox23,mitsuhashi23}). This is one of the reasons why redshift
completeness for systematically selected samples of DSFGs is
important, in order to not miss rare $z$$>$5 specimens with
potentially unusual dust properties.

While rare, the DSFG population reaches enormous bolometric
luminosities, which are typically dominantly transmitted at long
wavelengths due to dust obscuration. They thus exhibit far infrared
luminosities of $L_{\rm FIR}$$\sim$0.4--3$\times$10$^{13}$\,\lsol,
indicative of starbursts in excess of $\sim$500\,\msol\,yr$^{-1}$,
which are sometimes further flux--boosted by up to an order of
magnitude or more by gravitational lensing (e.g.,
\citealt{hezaveh13,rawle14,spilker16,fudamoto17,riechers20a}). Due to
these intense starbursts, the dust in DSFGs is typically relatively
warm, but it has been subject to debate whether or not they are warmer
than other star-forming galaxy populations (since high dust optical
depths may cause the warmest dust to be hidden), whether or not their
dust temperatures $T_{\rm dust}$ evolve with redshift, or if there is
evidence for a general evolution toward warmer dust in star-forming
galaxies with increasing redshift, which goes beyond the heating
contribution from the increasingly warmer cosmic microwave background
(CMB; e.g.,
\citealt{magnelli14,schreiber18,dudzeviciute20,riechers20a}).

The high levels of activity in DSFGs, combined with their high
observable brightness, provides access to a suite of diagnostic tools
which are often not accessible beyond the local universe. Of
particular interest are diagnostic lines such as the high-rotational
levels of CO which hold information about the gas excitation
mechanisms associated with the starbursts (see \citealt{cw13} for a
review), and light hydrides as tracers of the diffuse gas associated
with infalling material or large-scale outflows. A particularly
promising tool in this regard are the ground-state rotational
levels\footnote{There are three such levels:\ OH$^+$
  $N_J$=1$_0$$\to$0$_1$, $N_J$=1$_2$$\to$0$_1$, and
  $N_J$=1$_1$$\to$0$_1$, with their strongest components at rest
  frequencies of 909.1588, 971.8053, and 1033.058\,GHz, respectively.}
of the highly reactive OH$^+$ molecular ion, which is thought to show
enhanced abundances in diffuse atomic gas subjected to high cosmic ray
fluxes, such as expected in gas flows bathed in the radiation fields
created by intense starbursts (e.g.,
\citealt{hollenbach12}). Consequently, OH$^+$ absorption has now been
detected in $>$20 high-$z$ massive starbursts (e.g.,
\citealt{riechers13b,riechers21a,riechers21b,indriolo18,berta21,butler21}),
which makes it possible to start investigating statistical properties.

As an extension of this study, we thus have targeted an additional
$\sim$70 high-$z$ massive starbursts in at least one ground-state
OH$^+$ transition, which also provided simultaneous coverage of the
\ico\ transition in all cases (D.~Riechers et al., in
preparation). This sample includes galaxies from the South Pole
Telescope (SPT) survey of DSFGs \citep{vieira10}. As a byproduct, this
study has allowed us to independently confirm the redshifts of three
sources in this survey which could still have been considered
ambiguous based on previous studies
\citep{weiss13,strandet16,reuter20}. We thus report on the outcome of
these observations here, in the interest of redshift completeness of
the sample. In addition, we failed to detect both targeted lines in
one source in the entire sample, which was also selected from the SPT
survey. Given the potentially unusual nature of this source, we here
provide an in-depth analysis of this source, including a range of
revised properties.

We present the sample, data, and calibration in
Section~\ref{sec:data}, before presenting the outcome of the
observations in Section~\ref{sec:results} and the deeper analysis in
Section~\ref{sec:analysis}. A brief summary and conclusions are
provided in Section~\ref{sec:conclusions}. We use a concordance, flat
$\Lambda$CDM cosmology throughout, with $H_0$=69.6\,\kms\,Mpc$^{-1}$,
$\Omega_{\rm M}$=0.286, and $\Omega_{\Lambda}$=0.714.


\begin{deluxetable*}{ l c c c c c c c c }

\tabletypesize{\scriptsize}
\tablecaption{ALMA Observations. \label{t1}}
\tablehead{
\colhead{SPT name} & \colhead{LSB/USB center} & \colhead{Band} & \colhead{$N_{\rm ant}$} & \colhead{$\theta_{\rm maj}$$\times$$\theta_{\rm min}$\tablenotemark{\scriptsize a}} & \colhead{observing dates} & \colhead{$t_{\rm  on}$} & \colhead{complex gain} & \colhead{bandpass/} \\[-3mm]
                   & \colhead{(GHz)}                 &                &                       &                                                                    &                           & \colhead{(min)}       & \colhead{calibrator} & \colhead{flux$^\dagger$ calib}}
\startdata
{\em ALMA/ACA (1.8\,hr)\tablenotemark{\scriptsize b}:} & & & & & & & & \\
2354--58 (J2354--5815) & 348.85/360.77 & 7 & 9  & 5\farcs4$\times$3\farcs5/6\farcs4$\times$3\farcs2 & 2022 March 31    & 22.2 & J2357--5311 & J2253+1608  \\
0150--59 (J0150--5924) & 257.45/272.82 & 6 & 9  & 7\farcs1$\times$4\farcs8/6\farcs7$\times$4\farcs5 & 2022 April 12    & 19.2 & J0210--5101 & J2253+1608  \\
0314--44 (J0314--4452) & 248.88/262.59 & 6 & 9  & 6\farcs2$\times$4\farcs9/5\farcs9$\times$4\farcs6 & 2022 July 29     & 21.2 & J0334--4008 & J0538--4405 \\
0452--50 (J0452--5018) & 331.55/343.45 & 7 & 10 & 5\farcs0$\times$3\farcs4/4\farcs8$\times$3\farcs4 & 2022 May 21      & 19.7 & J0515--4556 & J0538--4405 \\
         & 304.84/316.75 & 7 & 10 & 6\farcs3$\times$3\farcs3/6\farcs0$\times$3\farcs2 & 2023 December 17 & 25.2 & J0515--4556 & J0538--4405 \\
\tableline
{\em Archival data (2.6\,hr)\tablenotemark{\scriptsize b}:} & & & & & & & & \\
0452--50 & 254.08/267.93 & 6 & 10 & 6\farcs9$\times$4\farcs4/6\farcs5$\times$4\farcs2 & 2019 October 3 \& 13 & 58.8 & J0455--4615 & J0538--4405 \\
         & 162.56/174.56 & 5 & 10 & 10\farcs3$\times$6\farcs6/9\farcs6$\times$6\farcs3 & 2019 November 10     & 94.9 & J0455--4615 & J0538--4405 \\
\tableline
{\em Literature data (3\,min)\tablenotemark{\scriptsize b}:} & & & & & & & & \\
0452--50 & 97.29/109.29 & 3 & 15\tablenotemark{\scriptsize c} & 6\farcs5$\times$5\farcs6/5\farcs8$\times$5\farcs0 & 2011 November 19 & 3.0  & J0455--4615 & J0538--4405 \\
         &         &   &    &                                                   &                  &      &             & Mars$^\dagger$ \\
\enddata
\tablenotetext{a}{Synthesized beam size.}
\tablenotetext{b}{Total on source time $t_{\rm on}$.}
\tablenotetext{c}{Observations using the 12\,m antennas in the main array. All other cases are the ACA 7\,m antennas.}
${}$\\[-10mm]
\end{deluxetable*}

\section{Data} \label{sec:data}

\subsection{Sample Selection}

The four targets analyzed in this work are selected from the SPT
survey of bright DSFGs (\citealt{reuter20}, and references therein),
and from a larger survey of high-$z$ starbursts targeting the
OH$^+$($N_J$=1$_1$$\to$0$_1$) line for which the first results were
reported by \citet{riechers21b}. The first three targets are included
here with the main purpose of unambiguously confirming their
spectroscopic redshifts,\footnote{These sources were reported as
single-line CO detections in the survey by \cite{reuter20}, which
results in ambiguous redshift identifications in at least two
cases. The redshift of SPT\,2354--58 was previously determined based
on a \dco\ emission line and an absorption line identified as
OH$^+$($N_J$=1$_2$$\to$0$_1$). For lensed galaxies, there is a chance
that absorption lines can be associated with the foreground lensing
galaxies (or other interlopers). While unlikely, we here include this
source among those confirmed for completeness.} as a byproduct of the
OH$^+$ survey.

The fourth target, SPT\,0452--50 is included here, because the
observations targeting OH$^+$ called its redshift identification into
question, as described in more detail in the following. Despite being
only a single source, its analysis is of particular interest in the
context of discussions of $T_{\rm dust}$--$z$ relations, since any
such relations depend on the underlying galaxy selection function.
For dusty galaxies with well-sampled dust spectral energy
distributions, the uncertainties in parameters for individual sources,
in general, are relatively minor, such that even rare outliers can
have an impact on perceived relations. As shown in Figure 8 of
\citet{reuter20}, SPT\,0452--50 was found to be a far outlier in the
$T_{\rm dust}$--$L_{\rm FIR}$--$z$ parameter space (see also
\citealt{strandet16}), suggesting that it could be an unusually cold
massive starburst with a dust temperature of only $T_{\rm
  dust}$=21$\pm$2\,K (the rest of their sample all have $T_{\rm
  dust}$$>$30\,K). SPT\,0452--50 would also be an outlier in $T_{\rm
  dust}$ in other DSFG samples with well-measured dust spectral energy
distributions (SEDs). As an example, \citet{ismail23} report $T_{\rm
  dust}$ for 125 {\em Herschel}-selected DSFGs at $z$=1.4--5.4, which
at face value show individual values as low as
20.5$^{+2.8}_{-2.0}$\,K.  However, these values are determined based
on optically-thin SED fits, which the same authors find to
underpredict $T_{\rm dust}$ by 5--15\,K compared to general SED fits
that account for dust optical depth effects.\footnote{Fitting their
coldest source, HELMS35, with a general SED fit following the method
described below yields $T_{\rm dust}$=41.9$^{+1.7}_{-3.9}$\,K.} As
such, no sources as cold as SPT\,0452--50 appear to be found in
broader DSFG samples when accounting for differences in SED fitting
methods. Such sources may be rare in current samples, because some
selection techniques to identify high-$z$ massive starbursts may
actively select against such sources. Should they exist in
considerable quantities, they could substantially alter our
understanding of $T_{\rm dust}$--$z$ relations in the early
universe. As such, this source is analyzed here in greater detail.

\subsection{New ALMA/ACA Observations}

We observed the redshifted \ico\ and OH$^+$($N_J$=1$_1$$\to$0$_1$)
lines ($\nu_{\rm rest}$=1036.9124 and 1033.0582\,GHz) toward the four
galaxies in our sample in Bands 6 and 7 with the ALMA/ACA, using nine
or ten 7\,m antennas covering 8.9--48.9\,m baselines (see
Tab.~\ref{t1}). For two galaxies, the observing setups also covered
the OH$^+$($N_J$=1$_2$$\to$0$_1$) line ($\nu_{\rm
  rest}$=971.8053\,GHz). Observations were carried out in cycle 8
under good to excellent weather conditions between 2022 March 31 and
July 29 (project ID:\ 2021.2.00062.S; PI:\ Riechers), spending between
19 and 22\,min on source per target. We also observed a second, lower
Band 7 setup for SPT\,0452--50 with the ALMA/ACA for 25\,min in cycle
10 on 2023 December 17 (project ID:\ 2023.1.01481.S; PI:\ Riechers).
Radio quasars close to the sources in sky projection were observed for
complex gain, bandpass, and absolute flux calibration (see Table
\ref{t1}; the bandpass and flux calibrators were identical where not
listed separately). The absolute flux calibration is estimated to be
reliable to within $<$10\%.

The ACA correlator was set up with two spectral windows of 1.875\,GHz
bandwidth (dual polarization) each per sideband, at a sideband
separation of typically 8\,GHz. A spectral resolution of 31.25\,MHz at
a channel spacing of 15.625\,MHz was chosen for all observations to
reduce calibration overheads. Thus, neighboring channels in spectra
shown at full resolution are not independent.

All data were calibrated aided by the calibration pipeline in
\casa\ version 6.2.1 \citep{mcmullin07}, and manually imaged using the
CLEAN algorithm via the {\tt tclean} task with ``natural'' baseline
weighting, resulting in the synthesized beam sizes listed in
Table~\ref{t1}. This is with the exception of the second Band 7
observations, where \casa\ version 6.5.4 was used for calibration. The
2022 (2023) data for SPT\,2354--58, 0150--59, 0314--44, and 0452--50
reach rms noise levels of 8.5/9.3, 2.6/3.2, 3.7/4.4, and 6.0/6.4
(4.1/5.8) mJy\,beam$^{-1}$ per 15.625\,MHz channel in the lower/upper
sideband (LSB/USB), respectively.

\begin{deluxetable*}{ l c l c l c l c }

\tabletypesize{\scriptsize}
\tablecaption{Continuum fluxes. \label{t2}}
\tablehead{
\colhead{SPT\,2354--58} & & \colhead{SPT\,0150--59} & & \colhead{SPT\,0314--44} & & \colhead{SPT\,0452--50} & \\[-3mm]
\colhead{$\nu_{\rm obs}$ (GHz)} & \colhead{Flux (mJy)} & \colhead{$\nu_{\rm obs}$ (GHz)} & \colhead{Flux (mJy)} & \colhead{$\nu_{\rm obs}$ (GHz)} & \colhead{Flux (mJy)} & \colhead{$\nu_{\rm obs}$ (GHz)} & \colhead{Flux (mJy)} }
\startdata
349 & 65.6$\pm$2.3 & 257 & 20.28$\pm$0.67 & 249 & 58.2$\pm$1.1 & 97 & 0.71$\pm$0.17 \\
361 & 67.1$\pm$2.8 & 273 & 24.23$\pm$0.68 & 263 & 65.1$\pm$0.9 & 109 & 1.61$\pm$0.23 \\
 & & & & & & 163 & 6.11$\pm$0.18 \\
 & & & & & & 175 & 8.41$\pm$0.23 \\
 & & & & & & 254 & 23.73$\pm$0.38 \\
 & & & & & & 268 & 27.26$\pm$0.39 \\
 & & & & & & 305 & 37.17$\pm$0.81 \\
 & & & & & & 317 & 38.15$\pm$0.98 \\
 & & & & & & 332 & 41.24$\pm$0.84 \\
 & & & & & & 343 & 43.26$\pm$0.97 \\
\enddata
\end{deluxetable*}

\begin{figure*}[tbh!]
\epsscale{0.94}
\plotone{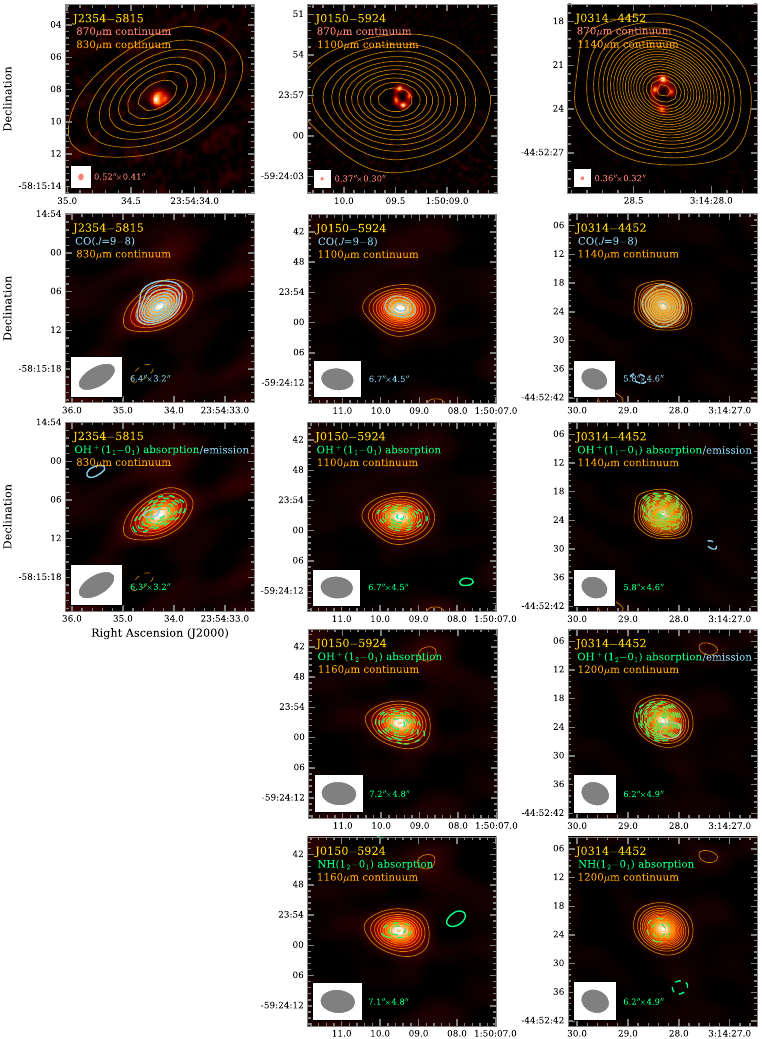}
\vspace{-2mm}

\caption{Line and continuum emission toward SPT\,2354--58, 0150--59,
  and 0314--44 (left to right). Top row:\ USB continuum contours
  overlaid on archival ALMA 870\,$\mu$m high-resolution continuum
  images (project IDs:\ 2011.0.00958.S, 2016.1.00231.S, and
  2019.1.01026.S, respectively; left background data previously
  published by \citealt{spilker16}). Second and third row:\ \ico\ and
  OH$^+$($N_J$=1$_1$$\to$0$_1$) emission/absorption contours overlaid
  on USB continuum emission and contours. Fourth and fifth
  row:\ OH$^+$ and NH $N_J$=1$_2$$\to$0$_1$ emission/absorption
  contours overlaid on LSB continuum emission and contours. Continuum
  contours (orange) are shown in steps of $\pm$3$\sigma$, where
  1$\sigma$=0.67 and 1.14 (2.3, 0.68, and 0.95) mJy\,beam$^{-1}$ in
  the LSB (USB; always ordered left to right). Line contours are shown
  in steps of 1$\sigma$, starting at $\pm$3$\sigma$, except for
  SPT\,0314--44, where \ico\ is shown in steps of 2$\sigma$. For the
  second row, 1$\sigma$=2.8, 1.1, and 1.3\,mJy\,beam$^{-1}$,
  respectively. For the third row, 1$\sigma$=3.8, 1.2, and 1.3 (2.3,
  N/A, and 1.5) mJy\,beam$^{-1}$ for absorption (emission),
  respectively. For the fourth row, 1$\sigma$=1.0 and 1.6 (N/A and
  1.7) mJy\,beam$^{-1}$ for absorption (emission), respectively. For
  the fifth row, 1$\sigma$=0.80 and 1.8\,mJy\,beam$^{-1}$,
  respectively. Negative contours are dashed. Observed-frame continuum
  wavelengths are indicated in each panel. The synthesized beam size
  is indicated in the bottom left corner of each panel.\label{f1}}
%
\end{figure*}

\begin{figure*}
\epsscale{1.18}
\plotone{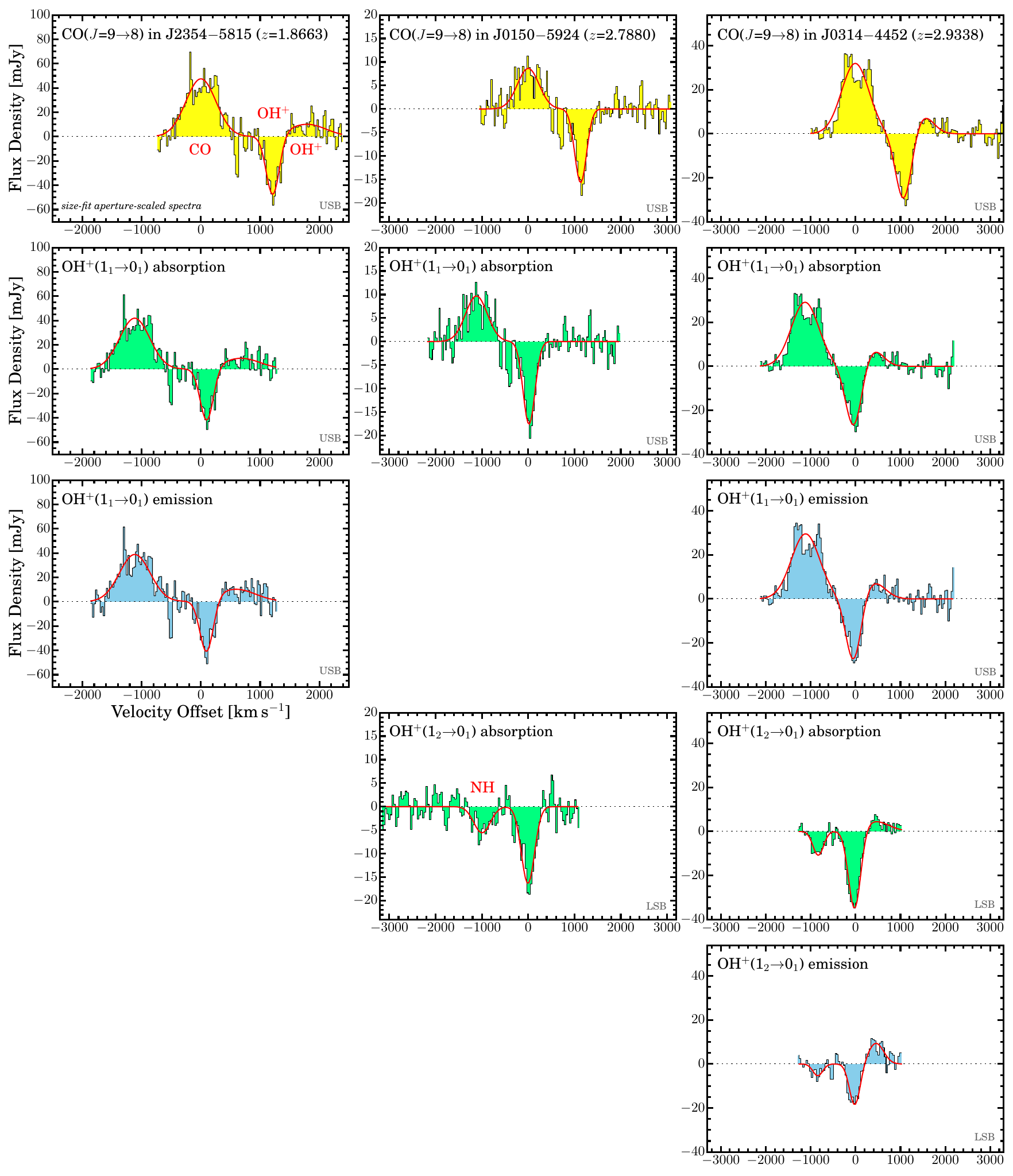}
\vspace{-5mm}

\caption{Line spectra toward SPT\,2354--58, 0150--59, and 0314--44 at
  a spectral resolution of 31.25\,MHz (histograms) and multi-component
  Gaussian fits to all features (red lines; left to right). Spectra
  are extracted from apertures scaled to the fitted size in the
  moment-0 map of the main line indicated in each panel after
  continuum subtraction, and scaled to the \ico\ systemic
  redshifts. Peak spectra are used where the size fits are most
  consistent with a point source. All features are fitted
  simultaneously, but only the parameters for the main features are
  adopted (no separate plots are shown for NH).\label{f2}}
%
\end{figure*}

\begin{figure}[h!]
\epsscale{1.0}
\plotone{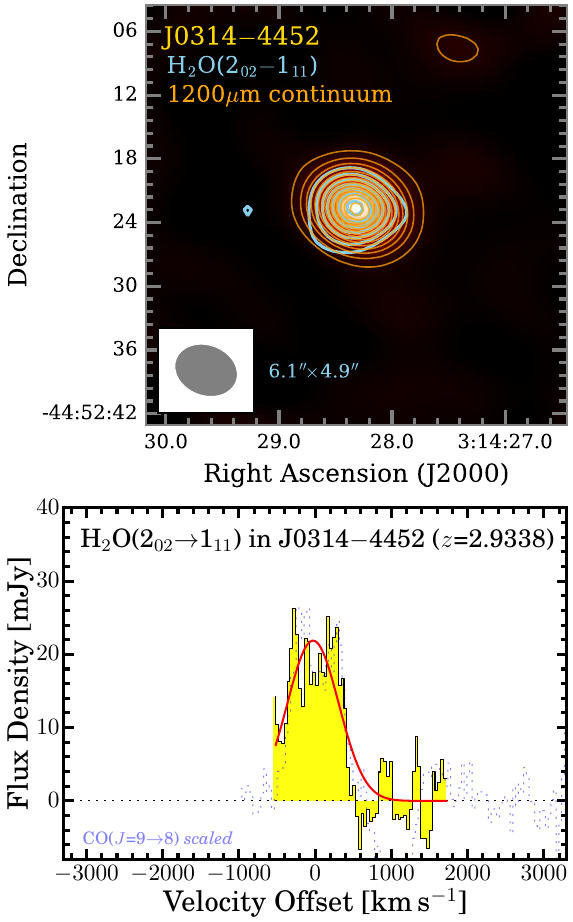}
\vspace{-2mm}

\caption{H$_2$O(2$_{02}$$\to$1$_{11}$) map (top) and spectrum (bottom)
  toward SPT\,0314--44, in the same style as in Figs.~\ref{f1} and
  \ref{f2}. For comparison, a scaled version of the \ico\ spectrum is
  shown as a dotted histogram (bottom).\label{f3}}
%
\end{figure}

\subsection{Archival ALMA/ACA Observations}

We also include archival cycle 7 ALMA/ACA Band 5 and 6 observations
targeting the two [CI] fine structure lines in SPT\,0452--50 at the
redshift of $z$=2.011 reported by \citet{reuter20} during two
observing runs each (project ID:\ 2019.1.00297.S; PI:\ Bethermin). The
details of these observations are provided in Table~\ref{t1}. The
correlator setups are the same as those in the previous subsection,
except for the tuning frequencies, and that a two times higher
spectral resolution (i.e., 7.8125\,MHz) was chosen for the spectral
windows expected to contain the lines (one in Band 5, two in Band
6). All data were calibrated aided by the calibration pipeline in
\casa\ version 6.2.1, and manually imaged. The Band 5 (Band 6) data
reach rms noise levels of 1.5/2.4 (4.4/4.1) mJy\,beam$^{-1}$ per
15.625\,MHz channel in the LSB/USB, respectively.

\subsection{Literature ALMA Observations}

For our analysis, we have also re-calibrated one of the cycle 0 ALMA
12\,m Band 3 observations of SPT\,0452--50 previously published by
(\citealt{weiss13}; project ID:\ 2011.0.00957.S; PI:\ Wei\ss ). These
observations were part of the line scan observations used to determine
its redshift, and reported to contain no line emission. The details of
these observations are provided in Table~\ref{t1}. The correlator
setups are the same as those in the previous subsection, except for
the tuning frequencies, and that a spectral resolution of 488.281\,kHz
was utilized. In addition, due to the poorly known flux scales of many
flux calibrators at the time, Mars was observed for flux
calibration. The data and calibration procedures were edited to be
compatible with current versions of \casa, and to reduce the excessive
edge channel flagging applied by default versions of the
calibration. All data were then calibrated aided by the calibration
pipeline in \casa\ version 6.2.1, and manually imaged, achieving an
rms noise level of 2.6\,mJy\,beam$^{-1}$ per 15.625\,MHz channel in
the lower sideband.

\section{Results} \label{sec:results}

\begin{deluxetable}{ l c c c }

\tabletypesize{\scriptsize}
\tablecaption{Line properties of the ``confirmed'' sources. \label{t3}}
\tablehead{
\colhead{} & \colhead{SPT\,2354--58} & \colhead{SPT\,0150--59} & \colhead{SPT\,0314--44}}
\startdata
$z_{\rm CO}$ & 1.8663$\pm$0.0002 & 2.7880$\pm$0.0004 & 2.9338$\pm$0.0002 \\
$I_{\rm CO(9-8)}$ (Jy\,\kms ) & 30.2$\pm$1.9 & 5.05$\pm$0.59 & 24.6$\pm$1.2 \\
d$v_{\rm FWHM}$ (\kms ) & 600$\pm$40 & 550$\pm$70 & 730$\pm$40 \\
$L'_{\rm CO(9-8)}$ (10$^{10}$\,$L_l$)\tablenotemark{\scriptsize a} & 6.72$\pm$0.42 & 2.27$\pm$ 0.27 & 12.04$\pm$0.59 \\
$L_{\rm CO(9-8)}$ (10$^8$\,\lsol )\tablenotemark{\scriptsize a} & 24.0$\pm$1.5 & 8.09$\pm$0.95 & 42.9$\pm$2.1 \\
\tableline
$I_{\rm H2O(202-111)}$ (Jy\,\kms ) & & & 18.4$\pm$1.5 \\
d$v_{\rm FWHM}$ (\kms ) & & & 795$\pm$75 \\
$v_0-v_{\rm 0,CO}$ (\kms ) & & & --28$\pm$29 \\
$L'_{\rm H2O}$ (10$^{10}$\,$L_l$)\tablenotemark{\scriptsize a} & & & 9.9$\pm$0.8 \\
$L_{\rm H2O}$ (10$^8$\,\lsol )\tablenotemark{\scriptsize a} & & & 30.6$\pm$2.5 \\
\tableline
$I_{\rm OH+(11-01)}^{\rm abs}$ (Jy\,\kms ) & --12.2$\pm$2.8 & --5.6$\pm$0.9 & --12.3$\pm$4.3 \\
d$v_{\rm FWHM}$ (\kms ) & 260$\pm$40 & 300$\pm$30 & 420$\pm$70 \\
$v_0-v_{\rm 0,CO}$ (\kms ) & 100$\pm$13 & 18$\pm$13 & --40$\pm$50 \\
$\tau_{\rm OH+}$ & 0.86$^{+0.38}_{-0.27}$ & 0.42$^{+0.09}_{-0.08}$ & 0.42$^{+0.20}_{-0.08}$ \\
$\tau_{\rm OH+}$d$v$ (\kms ) & 380$^{+170}_{-120}$ & 271$^{+57}_{-52}$ & 300$^{+140}_{-55}$ \\
$N({\rm OH^+})$ (10$^{14}$\,cm$^{-2}$)\tablenotemark{\scriptsize b} & 18.5$^{+8.3}_{-5.8}$ & 13.2$^{+2.8}_{-2.5}$ & 14.6$^{+6.8}_{-2.7}$ \\
$N({\rm H})$ (10$^{22}$\,cm$^{-2}$)\tablenotemark{\scriptsize b} & 11.7$^{+5.2}_{-3.7}$ & 8.3$^{+1.7}_{-1.6}$ & 9.2$^{+4.3}_{-1.7}$ \\
\tableline
$I_{\rm OH+(12-01)}^{\rm abs}$ (Jy\,\kms ) & & --5.8$\pm$0.9 & --12.4$\pm$1.8 \\
d$v_{\rm FWHM}$ (\kms ) & & 330$\pm$30 & 320$\pm$20 \\
$v_0-v_{\rm 0,CO}$ (\kms ) & & --6$\pm$13 & --20$\pm$7 \\
$\tau_{\rm OH+}$ & & 0.50$^{+0.12}_{-0.11}$ & 0.57$^{+0.29}_{-0.11}$ \\
$\tau_{\rm OH+}$d$v$ (\kms ) & & 274$^{+65}_{-58}$ & 320$^{+160}_{-60}$ \\
\tableline
$I_{\rm OH+(11-01)}^{\rm em}$ (Jy\,\kms ) & 8.6$\pm$4.1 & & 4.8$\pm$3.2 \\
d$v_{\rm FWHM}$ (\kms ) & 800$\pm$500 & & 600$\pm$500 \\
$v_0-v_{\rm 0,CO}$ (\kms ) & 600$\pm$200 & & $<$400$\pm$300 \\
$L'_{\rm OH+(1-0)}$ (10$^{10}$\,$L_l$)\tablenotemark{\scriptsize a} & 1.93$\pm$0.92 & & 2.4$\pm$1.6 \\
$L_{\rm OH+(1-0)}$ (10$^8$\,\lsol )\tablenotemark{\scriptsize a} & 6.8$\pm$3.2 & & 8.4$\pm$5.6 \\
\tableline
$I_{\rm OH+(12-01)}^{\rm em}$ (Jy\,\kms ) & & & 4.0$\pm$1.4 \\
d$v_{\rm FWHM}$ (\kms ) & & & 400$\pm$90 \\
$v_0-v_{\rm 0,CO}$ (\kms ) & & & 460$\pm$40 \\
$L'_{\rm OH+(1-0)}$ (10$^{10}$\,$L_l$)\tablenotemark{\scriptsize a} & & & 2.2$\pm$0.8 \\
$L_{\rm OH+(1-0)}$ (10$^8$\,\lsol )\tablenotemark{\scriptsize a} & & & 6.5$\pm$2.3 \\
\tableline
$I_{\rm NH(12-01)}^{\rm abs}$ (Jy\,\kms ) & & --1.91$\pm$0.75 & --3.57$\pm$0.76 \\
d$v_{\rm FWHM}$ (\kms ) & & 420$\pm$100 & 280$\pm$40 \\
$v_0-v_{\rm 0,CO}$ (\kms ) & & --170$\pm$40 & --6$\pm$15 \\
$\tau_{\rm NH}$ & & 0.20$^{+0.09}_{-0.08}$ & 0.19$^{+0.04}_{-0.04}$ \\
$\tau_{\rm NH}$d$v$ (\kms ) & & 122$^{+56}_{-51}$ & 71$^{+17}_{-16}$ \\
\enddata
\tablenotetext{a}{Given in units of $L_l$=K\,\kms\,pc$^2$. Luminosities are ``apparent'', i.e., not corrected for gravitational magnification ($\mu_{\rm L}$=6.3$\pm$0.4 for SPT\,2354--58; \citealt{spilker16}; unknown for the other sources).}
\tablenotetext{b}{See \citep{indriolo18,bialy19,riechers21b} for conversions from optical depth to column densities $N$ of OH$^+$ and atomic hydrogen.}
${}$\\[-12.5mm]
\end{deluxetable}

\subsection{Confirmed Sources}

We successfully detect rest-frame 290\,$\mu$m continuum emission
toward SPT\,2354--58, 0150--59, and 0314--44 (Fig.~\ref{f1}). The peak
significance of the detections in the lower (upper) sidebands is 31,
30, and 47$\sigma$ (23, 36, and 64$\sigma$), respectively. The
emission is marginally resolved at best and smaller than the beam size
in all cases, consistent with the modest sizes of the strongly-lensed
galaxies in high resolution ALMA images. Continuum fluxes were
extracted from two-dimensional Gaussian fitting in the image
plane.\footnote{Although the sources are substantially more compact
than the beam sizes, residual phase errors may result in finite fitted
sizes, such that peak fluxes would underestimate the total fluxes. As
such, it is best practice to extract fluxes from size fits if the
signal-to-noise ratio of the detection is sufficiently high, as is the
case for all continuum detections reported here.} All continuum fluxes
are reported in Table~\ref{t2}. To investigate the presence of
emission and absorption lines, continuum emission was subtracted from
all data cubes in the visibility plane, excluding spectral regions of
potential lines in the fitting process. The subtraction accounts for
the spectral slope of the continuum where measurable within the
line-free spectral ranges.

\begin{figure*}
\epsscale{1.15}
\plotone{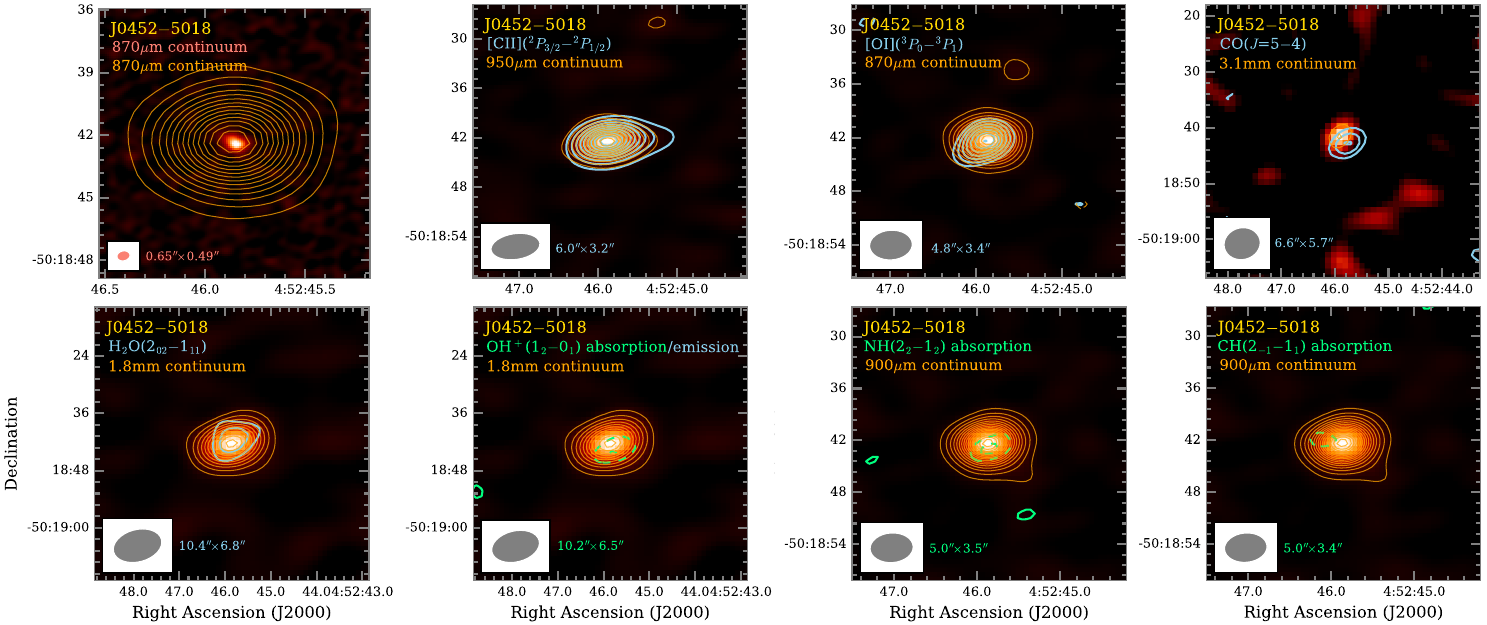}
\vspace{-2mm}

\caption{Line and continuum emission toward SPT\,0452--50. Top
  left:\ USB continuum contours from the [OI] observations, overlaid
  on an archival ALMA 870\,$\mu$m high-resolution continuum image
  (project ID:\ 2011.0.00958.S; background data previously published
  by \citealt{spilker16}). Remainder top row, left to right:\ [CII]
  158\,$\mu$m, [OI] 146\,$\mu$m, and \eco\ contours overlaid on the
  continuum emission at the same wavelengths, in the same style as
  Fig.~\ref{f1}. Bottom row, left to
  right:\ H$_2$O(2$_{02}$$\to$1$_{11}$) emission contours and
  OH$^+$($N_J$=1$_1$$\to$0$_1$), NH($N_J$=2$_2$$\to$1$_2$) and
  CH($N_J$=2$_{-1}$$\to$1$_1$) absorption contours overlaid on the
  continuum emission at the same wavelengths. Continuum contours in
  the second to eighth panels are shown in steps of $\pm$3$\sigma$,
  where 1$\sigma$=0.98, 0.97, 0.17, 0.18, 0.18, 0.84, and
  0.84\,mJy\,beam$^{-1}$ (except for CO $J$=5$\to$4, where contours
  start at $\pm$3$\sigma$ and are shown in steps of 1$\sigma$),
  respectively. Line contours are shown in steps of 1$\sigma$,
  starting at $\pm$3$\sigma$, except for [CII], where contours are
  shown in steps of 3$\sigma$. In the same order, 1$\sigma$=1.9, 1.6,
  0.69, 0.42, 0.68, 3.6, and 3.6\,mJy\,beam$^{-1}$,
  respectively. OH$^+$ emission contours
  (1$\sigma$=0.89\,mJy\,beam$^{-1}$) are also included but do not
  appear because the signal is $<$3$\sigma$ significant.\label{f4}}
%
\end{figure*}

We detect \ico\ emission at 9.5, 5.4, and 16$\sigma$ peak significance
and OH$^+$($N_J$=1$_1$$\to$0$_1$) absorption at 7.9, 7.2, and
16$\sigma$ peak significance toward SPT\,2354--58, 0150--59, and
0314--44, respectively (Fig.~\ref{f1}). We also detect
OH$^+$($N_J$=1$_2$$\to$0$_1$) absorption at 7.7 and 14$\sigma$ peak
significance and NH($N_J$=1$_2$$\to$0$_1$) absorption at 4.6 and
4.6$\sigma$ peak significance toward SPT\,0150--59 and 0314--44,
respectively.  The emission components of
OH$^+$($N_J$=1$_1$$\to$0$_1$) are tentatively detected at 3.3, and
3.7$\sigma$ peak significance toward SPT\,2354--58 and 0314--44,
respectively. The emission component of OH$^+$($N_J$=1$_2$$\to$0$_1$)
is also tentatively detected at 3.9$\sigma$ peak significance toward
SPT\,0314--44. Those OH$^+$ lines seen in emission are detected as
inverse P-Cygni profiles, which is consistent with what is expected
for outflowing gas (Fig.~\ref{f2} and Table~\ref{t3}).\footnote{Line
fluxes were extracted from the moment-0 maps in the same way as
continuum fluxes when the signal-to-noise ratio was sufficient, and
then compared to peak fluxes to evaluate whether or not a line is
resolved in the data and thus requires an aperture correction to its
peak flux spectrum as extracted from the brightest pixel in the
moment-0 map. For lines too faint to provide reliable size fits,
aperture corrections were adopted from the brightest emission line for
emission lines, and from the continuum for absorption lines, but the
center of the aperture was placed on the peak pixel of the faint line
in its moment-0 map. This is to allow for small spatial offsets, e.g.,
between the absorption and emission components of lines with P-Cygni
profiles. For reference, the average aperture correction in
Fig.~\ref{f2} is 13.3\%, with a factor of $\sim$2 scatter toward the
most extreme values.}  In SPT\,2354--58, the OH$^+$ absorption is
slightly redshifted, while in the other two galaxies, the centroid
velocities are consistent with systemic within $<$1--2$\sigma$. Given
the broad CO line widths compared to the shifts, the absorption
components alone thus do not conclusively show that the gas is
outflowing. In both SPT\,2354--58 and 0314--44, the emission
components are far redshifted -- which is consistent with outflowing
gas in these two sources.\footnote{While differential lensing between
different gas components could partially explain differences in line
shapes between different species, we expect these effects to be
subdominant to real differences in the velocities of different gas
components. While a broader analysis is beyond the scope of this work,
these considerations are supported by the analysis of high-resolution
CO and OH$^+$ observations of other, similarly bright DSFGs (e.g.,
\citealt{berta21,butler21}).}  Lastly, we also detect
H$_2$O($J_{K_aK_c}$=2$_{02}$$\to$1$_{11}$) emission at 15$\sigma$ peak
significance toward SPT\,0314--44 (Fig.~\ref{f3}). This unambiguously
confirms the redshifts of all sources. Line parameters were obtained
from Gaussian fitting to the line profiles, using multiple Gaussian
components where appropriate. From the \ico\ lines, we measure
systemic $z_{\rm CO}$=1.8663$\pm$0.0002, 2.7880$\pm$0.0004, and
2.9338$\pm$0.0002, respectively. All line parameters are summarized in
Table~\ref{t3}. These observations will be analyzed further in concert
with the full sample of $\sim$70 sources observed in the same program
(D.\ Riechers et al., in prep.).\footnote{In parallel to this work,
\citet{gururajan23} have also independently confirmed the redshifts of
SPT\,2354--58 and 0150--59 based on \gco\ and [CI] detections.}


\begin{deluxetable*}{ l c c c c c c c }

\tabletypesize{\scriptsize}
\tablecaption{Line properties of SPT\,0452--50. \label{t4}}
\tablehead{
  Line & \colhead{$I_{\rm line}$} & \colhead{d$v_{\rm FWHM}$} & \colhead{$v_0-v_{\rm 0,CII}$} & \colhead{$L'_{\rm line}$\tablenotemark{\scriptsize a}}   & \colhead{$L_{\rm line}$\tablenotemark{\scriptsize a}} & \colhead{$\tau_{\rm line}$} & \colhead{$\tau_{\rm line}$d$v$} \\[-3mm]
       & \colhead{(Jy\,\kms )} & \colhead{(\kms )}         & \colhead{(\kms )}            & \colhead{(10$^{10}$\,$L_l$)} & \colhead{(10$^8$\,\lsol )} & \colhead{} & \colhead{(\kms )}   
}
\startdata
\bco\tablenotemark{\scriptsize \scriptsize b} & 0.96$\pm$0.12 & 610$\pm$60  &             & 22.3$\pm$2.8 & 0.88$\pm$0.11 & & \\
\eco & 2.80$\pm$0.67 & 640$\pm$180 & --70$\pm$75 & 10.4$\pm$2.5 & 6.4$\pm$1.5 & & \\
\nco & $<$1.3 & {\em (630)} & & $<$0.64 & $<$8.5 & & \\
\pco & $<$2.7 & {\em (630)} & & $<$1.0 & $<$20 & & \\
\rco & $<$2.0 & {\em (630)} & & $<$0.57 & $<$16 & & \\
H$_2$O($J_{K_aK_c}$=2$_{02}$$\to$1$_{11}$)  & 2.02$\pm$0.52 & 860$\pm$260 & 150$\pm$100 & 2.56$\pm$0.66 & 7.9$\pm$2.0 & & \\
H$_2$O($J_{K_aK_c}$=4$_{13}$$\to$3$_{22}$)  & $<$1.4 & {\em (630)} & & $<$0.41 & $<$12 & & \\
H$_2$O($J_{K_aK_c}$=4$_{13}$$\to$4$_{04}$)  & $<$1.5 & {\em (630)} & & $<$0.73 & $<$9.6 & & \\
\oh & $<$4.1 & {\em (1130)} & & $<$1.5 & $<$30 & & \\
OH$^+$($N_J$=1$_2$$\to$0$_1$) abs. &   --0.89$\pm$0.47 & 280$\pm$90 & --150$\pm$40 & & & 0.82$^{+1.10}_{-0.51}$ & 300$^{+410}_{-190}$ \\
OH$^+$($N_J$=1$_2$$\to$0$_1$) em.  & {\em 0.5$\pm$0.2} & {\em 160$\pm$60} & {\em 595$\pm$25} & {\em 0.65$\pm$0.26} & {\em 1.9$\pm$0.8} & & \\
NH($N_J$=2$_2$$\to$1$_2$)     & $\sim$(--2.94$\pm$0.72)    & (200) & & & & $\sim$(0.50$^{+0.17}_{-0.15}$) & $>$(100$^{+35}_{-30}$)\\
CH($N_J$=2$_{-1}$$\to$1$_1$)   & {\em $<$(--1.78$\pm$0.51)} & {\em (140)} & & & & {\em 0.93$^{+0.58}_{-0.36}$} & {\em $>$(130$^{+80}_{-50}$)} \\
\cii & 54.8$\pm$4.0 & 630$\pm$50  & 0 & 18.8$\pm$1.4 & 410$\pm$30 & & \\
\oi  & 7.4$\pm$1.3 & 600$\pm$120 & 90$\pm$50 & 2.15$\pm$0.38 & 60$\pm$11 & & \\
\enddata
\tablenotetext{a}{Given in units of $L_l$=K\,\kms\,pc$^2$. Luminosities are ``apparent'', i.e., not corrected for gravitational magnification ($\mu_{\rm L}$=1.7$\pm$0.1; \citealt{spilker16}).}
\tablenotetext{b}{Line flux and width adopted from \citet{aravena16c}.}
\tablecomments{All quoted upper limits are 3$\sigma$.}
\end{deluxetable*}

\subsection{Revised Redshift}

We successfully detect continuum emission in the Band 5, 6, and 7
observations of SPT\,0452--50 presented here for the first time
(Fig.~\ref{f4}). The peak significance of the detections ranges
between 34$\sigma$ and 70$\sigma$ (Table~\ref{t2}) and remains
virtually unresolved, as is expected based on its compact size seen at
higher spatial resolution (Fig.~\ref{f4}, top left).

\begin{figure*}
\epsscale{1.18}
\plotone{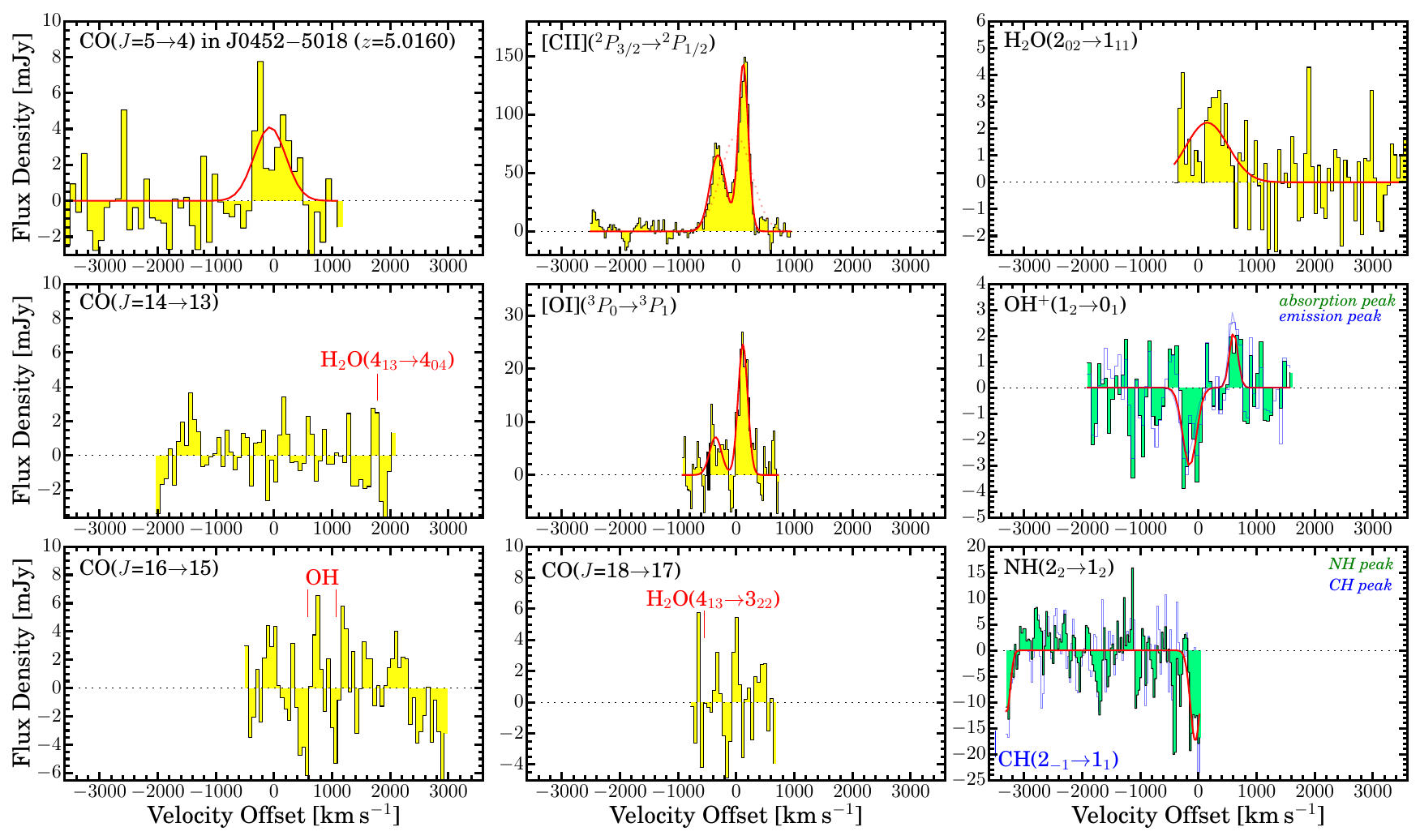}
\vspace{-5mm}

\caption{Line spectra toward SPT\,0452--50 a spectral resolution of
  31.25\,MHz (histograms; except CO upper limit spectra which are
  shown at half the resolution) and multi-component Gaussian fits to
  all features, in the same style as Fig.~\ref{f2}. Spectra are scaled
  to the [CII] flux-weighted systemic redshift (center velocity of
  single-Gaussian fit; dotted curve). Red/blue markers indicate the
  center velocities of additional features. The OH$^+$ emission and CH
  absorption peaks are tentative and require independent
  confirmation. Fit parameters to the NH/CH spectra are unreliable due
  to the incomplete line coverage, and thus, are discarded in the
  analysis.\label{f5}}
%
\end{figure*}

We however do not detect the targeted \ico\ and OH$^+$($N_{\rm
  J}$=1$_1$$\to$0$_1$) lines at $z$=2.011. Instead, the data show a
single emission line detected at 9.0$\sigma$ peak significance at an
unexpected frequency (Fig.~\ref{f5}). This line would be consistent
with the expected redshifted frequency of [OI] 146\,$\mu$m if the two
CO lines reported by \citet{aravena16c} and \citet{weiss13} were to be
due to \bco\ and \fco\ emission at $z$=5.016, instead of \aco\ and
\cco\ emission at $z$=2.011 as previously reported. However, this
redshift was seemingly ruled out due to the non-detection of \eco\ in
the 3\,mm line scan reported by \citet{weiss13}.

To further investigate this issue, we analyzed archival data targeting
both [CI] lines in this source at $z$=2.011. No [CI] or CO lines are
detected in these observations, but they show weak evidence for a
H$_2$O 2$_{02}$$\to$1$_{11}$ line at 4.8$\sigma$ significance at a
frequency consistent with $z$=5.016. We then re-analyzed the 3\,mm
line scan data reported by \citet{weiss13}. These data show weak
evidence for a \eco\ line at 5.1$\sigma$ significance at a frequency
consistent with $z$=5.016. We speculate that this line may have been
missed due to the conservative edge channel flagging applied by the
standard ALMA calibration procedures, but we cannot confirm this based
on the archived data products. This line, however, is much weaker than
the \cco\ line reported by the same authors (which, in this scenario,
would be CO $J$=6$\to$5), resulting in unrealistically high line
brightness temperature ratios.

Given the reliance on weak line features and unusual circumstances, we
targeted the [CII] line at $z$=5.016 to break the degeneracy between
the different scenarios. We detect [CII] emission at 29$\sigma$ peak
significance, which unambiguously determines the redshift to be
$z_{\rm CII}$=5.0160$\pm$0.0005 (Fig.~\ref{f4}). This implies that the
line detected by \citet{aravena16c} is \bco\ instead of \aco, and the
line detected by \citet{weiss13}, if confirmed, is \fco\ instead of
\cco.

Upon closer inspection, we also find evidence for three weak
absorption features due to OH$^+$($N_J$=1$_1$$\to$0$_1$),
NH($N_J$=2$_2$$\to$1$_2$) and CH($N_J$=2$_{-1}$$\to$1$_1$) at
4.0$\sigma$, 4.3$\sigma$, and 3.4$\sigma$ peak significance. OH$^+$
also shows evidence for a redshifted emission component, but only at
2.5$\sigma$ peak significance. Together with the blueshift of the
absorption component, this suggests the presence of outflowing gas.
NH and CH are only partially covered by the bandpass, such that their
line optical depths remain uncertain. All line parameters are
summarized in Table~\ref{t4}.

\section{Analysis and Discussion} \label{sec:analysis}

\subsection{Spectral Energy Distribution Modeling}

Motivated by the added dust photometry for the first three sources and
the revised redshift and additional data for the fourth source, we
have re-fit their SEDs based on modified black body (MBB) models to
their rest-frame far-infrared to millimeter wavelength emission. As
described in more detail in our previous works (e.g.,
\citealt{riechers13b,dowell14}), we use a Markov Chain Monte Carlo
based approach for this purpose. The procedure, called
\texttt{MBB$\_$EMCEE}, uses the dust temperature $T_{\rm dust}$, the
spectral slope of the dust emissivity $\beta_{\rm IR}$, and the
wavelength $\lambda_0$ where the dust optical depth reaches unity as
the main fit parameters. To approximately capture the falloff of the
SEDs on the far Wien side, we join the MBB to a power law with the
shape $\propto$$\nu^\alpha$ on the short wavelength side of its
peak. All fits are normalized to the observed-frame 500\,$\mu$m
continuum flux of each source. As done in our previous work, we place
a broad prior of $\beta_{\rm IR}$=1.8$\pm$0.6 on the dust emissivity
parameter to guide the fitting. This choice is motivated by the
$\beta_{\rm IR}$ found for molecular clouds in the Milky Way (e.g.,
\citealt{planck11}).

\begin{figure*}
\epsscale{1.16}
\plotone{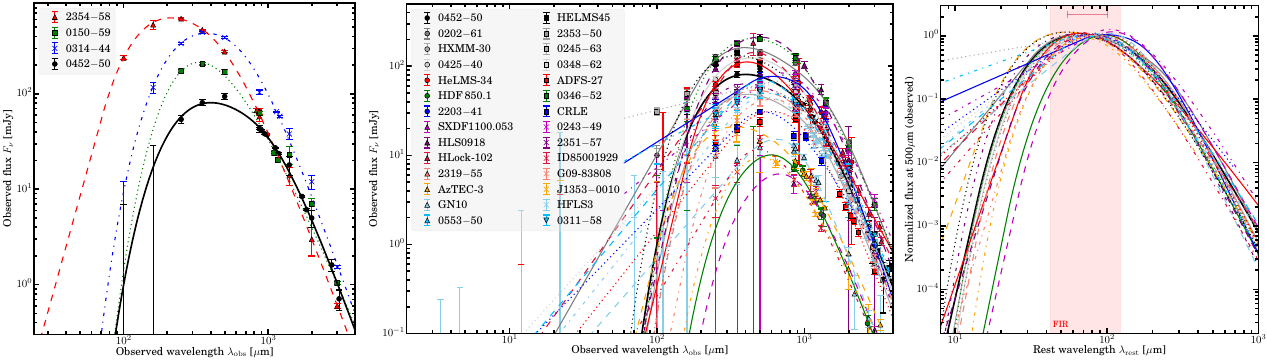}
\vspace{-2mm}

\caption{Spectral energy distributions of the ``confirmed'' sample
  (left; symbols) and $z$$>$5 DSFGs (middle/right) and MBB SED fits
  (lines; Tables~\ref{t5} and \ref{t6}) to the
  data. Left/middle:\ Observed-frame apparent fluxes. SPT\,0452--50 is
  shown in both panels for reference. Right:\ 500\,$\mu$m--normalized
  rest-frame models ($z$$>$5 sample only). The shaded region indicates
  the FIR luminosity range. The top bar indicates the range in peak
  wavelengths. \label{f6}}
%
\end{figure*}

We have also re-fit the dust SEDs of all known $z$$>$5 DSFGs with the
same procedure, except in those cases where this had already been done
in the literature (\citealt{riechers20a,riechers21a,riechers21b}). The
main motivation for this step were the significant changes in
parameters reported for the SPT sample between the procedures used by
\citet{strandet16} and \citet{reuter20}. While different approaches to
the fitting are valid, we decided to restrict our comparison to
samples fit with our procedure in order to minimize potential biases
in the comparison. That being said, the differences between parameters
reported by us and \citet{reuter20} for the same sources are mostly
minor, and can perhaps be explained by the inclusion of additional
photometry in our fits. The results for the galaxies analyzed in
detail in this work are summarized in Table~\ref{t5}, and those for
the entire $z$$>$5 DSFG sample are summarized in Table~\ref{t6}. The
corresponding data and fits are shown in Fig.~\ref{f6}.

\begin{deluxetable}{ l c c c c }

\tabletypesize{\scriptsize}
\tablecaption{SED parameters for the target sample. \label{t5}}
\tablehead{
& \colhead{SPT\,2354--58} & \colhead{SPT\,0150--59} & \colhead{SPT\,0314--44} & \colhead{SPT\,0452--50}}
\startdata
$T_{\rm dust}$ (K) & 69.2$^{+0.8}_{-1.2}$ & 45.9$^{+4.3}_{-8.5}$ & 54.9$^{+0.4}_{-2.3}$ & 76.2$^{+2.5}_{-2.5}$ \\
$\beta_{\rm IR}$ & 2.18$^{+0.03}_{-0.09}$ & 1.73$^{+0.02}_{-0.09}$ & 2.44$^{+0.01}_{-0.10}$ & 2.30$^{+0.13}_{-0.13}$ \\
$\lambda_{\rm peak}$ ($\mu$m) & 74$^{+1}_{-1}$ & 86$^{+3}_{-1}$ & 93$^{+3}_{-1}$ & 67$^{+2}_{-2}$ \\
$\lambda_0$ ($\mu$m) & 232$^{+4}_{-9}$ & 108$^{+37}_{-61}$ & 233$^{+2}_{-16}$ & 229$^{+9}_{-9}$ \\
$L_{\rm FIR}$ (10$^{13}$\,\lsol )\tablenotemark{\scriptsize a} & 5.82$^{+0.11}_{-0.08}$ & 3.25$^{+0.13}_{-0.19}$ & 7.44$^{+0.19}_{-0.15}$ & 4.01$^{+0.15}_{-0.14}$ \\
$L_{\rm IR}$ (10$^{13}$\,\lsol )\tablenotemark{\scriptsize a} & 9.40$^{+0.22}_{-0.33}$ & 5.00$^{+0.09}_{-0.99}$ & 10.76$^{+0.27}_{-0.35}$ & 6.96$^{+0.42}_{-0.46}$ \\
\enddata
\tablenotetext{a}{Apparent values not corrected for gravitational magnification ($\mu_{\rm L}$=6.3$\pm$0.4 and 1.7$\pm$0.1 for SPT\,2354--58 and 0452--50, respectively; \citealt{spilker16}). $L_{\rm FIR}$ ($L_{\rm IR}$) is integrated over 42.5--122.5\,$\mu$m (8--1000\,$\mu$m) in the rest frame.}
${}$\\[-12.5mm]
\end{deluxetable}

\begin{deluxetable*}{ l c c c c c c c c }

\tabletypesize{\scriptsize}
\tablecaption{SED parameters for all known $z$$>$5 DSFGs. \label{t6}}
\tablehead{
  \colhead{Name} & \colhead{$z$} & \colhead{$\mu_{\rm L}$\tablenotemark{\scriptsize a}} & \colhead{$T_{\rm dust}$} & \colhead{$\beta_{\rm IR}$} & \colhead{$\lambda_{\rm peak}$} & \colhead{$\lambda_0$} & \colhead{$L_{\rm FIR}$\tablenotemark{\scriptsize b}} & \colhead{$L_{\rm IR}$\tablenotemark{\scriptsize b}} \\[-3mm]
  & & & \colhead{(K)} & & \colhead{($\mu$m)} & \colhead{($\mu$m)} & \colhead{(10$^{13}$\lsol )} & \colhead{(10$^{13}$\lsol )}
 }
\startdata
SPT\,0452--50 & 5.0160 & 1.7$\pm$0.1   & 76.2$^{+2.5}_{-2.5}$ & 2.30$^{+0.13}_{-0.13}$ & 67$^{+2}_{-2}$ & 229$^{+9}_{-9}$ & 4.01$^{+0.15}_{-0.14}$ & 6.96$^{+0.42}_{-0.46}$ \\
SPT\,0202--61 & 5.0182 & 8.3$\pm$1.0   & 79.1$^{+2.5}_{-2.6}$  & 2.74$^{+0.23}_{-0.21}$ & 64$^{+2}_{-2}$ & 297$^{+17}_{-15}$ & 8.04$^{+0.25}_{-0.21}$ & 15.11$^{+0.90}_{-0.84}$ \\
HXMM-30	      & 5.0940 & $s$           & 74.2$^{+4.7}_{-4.6}$  & 2.64$^{+0.37}_{-0.39}$ & 69$^{+4}_{-4}$ & 242$^{+29}_{-34}$ & 2.47$^{+0.18}_{-0.17}$ & 4.41$^{+0.42}_{-0.64}$ \\ 
SPT\,0425--40 & 5.1353 & $s$           & 75.7$^{+1.9}_{-5.7}$  & 2.72$^{+0.44}_{-0.39}$ & 68$^{+4}_{-3}$ & 211$^{+29}_{-41}$ & 6.11$^{+0.30}_{-0.20}$ & 10.62$^{+0.40}_{-0.86}$ \\
HeLMS-34      & 5.1614 & $s$           & 63.0$^{+7.8}_{-8.3}$  & 1.36$^{+0.14}_{-0.18}$ & 66$^{+2}_{-2}$ & 93$^{+50}_{-64}$ & 5.38$^{+0.21}_{-0.20}$ & 8.65$^{+0.51}_{-0.78}$ \\
HDF\,850.1    & 5.1833 & 1.6$\pm$0.3   & 51.6$^{+15.6}_{-15.1}$ & 2.67$^{+0.30}_{-0.30}$ & 85$^{+13}_{-13}$ & 136$^{+45}_{-49}$ & 0.52$^{+0.17}_{-0.18}$ & 0.74$^{+0.30}_{-0.29}$ \\
SPT\,2203--41 & 5.1937 & $s$           & 54.0$^{+6.1}_{-6.0}$  & 3.16$^{+0.34}_{-0.32}$ & 94$^{+9}_{-9}$ & 219$^{+27}_{-31}$ & 3.48$^{+0.21}_{-0.19}$ & 8.52$^{+2.16}_{-3.04}$ \\
SXDF1100.053  & 5.2383 &               & 61.6$^{+19.0}_{-20.7}$ & 2.58$^{+0.26}_{-0.26}$ & 83$^{+20}_{-20}$ & 208$^{+40}_{-69}$ & 0.40$^{+0.15}_{-0.17}$ & 0.62$^{+0.31}_{-0.32}$ \\
HLS0918	      & 5.2430 & 8.9$\pm$1.9   & 64.9$^{+2.5}_{-3.2}$  & 2.69$^{+0.30}_{-0.26}$ & 78$^{+3}_{-3}$ & 196$^{+4}_{-31}$ & 10.26$^{+0.59}_{-0.41}$ & 15.95$^{+0.64}_{-1.07}$ \\
HLock-102     & 5.2915 & 12.5$\pm$1.2  & 55.7$^{+5.1}_{-7.5}$  & 1.80$^{+0.07}_{-0.08}$ & 73$^{+2}_{-1}$ & 99$^{+31}_{-43}$ & 6.82$^{+0.15}_{-0.13}$ & 9.67$^{+0.34}_{-0.50}$ \\
SPT\,2319--55 & 5.2927 & 7.9$\pm$1.9   & 68.6$^{+4.0}_{-4.8}$  & 2.72$^{+0.35}_{-0.31}$ & 75$^{+5}_{-5}$ & 234$^{+23}_{-29}$ & 2.78$^{+0.21}_{-0.19}$ & 4.62$^{+0.43}_{-0.57}$ \\
AzTEC-3	      & 5.2980 &               & 92.5$^{+15.4}_{-15.9}$ & 2.09$^{+0.21}_{-0.21}$ & 55$^{+8}_{-8}$ & 181$^{+33}_{-34}$ & 1.12$^{+0.16}_{-0.16}$ & 2.55$^{+0.73}_{-0.74}$ \\
GN10	      & 5.3031 &               & 48.8$^{+9.0}_{-11.2}$ & 3.18$^{+0.26}_{-0.21}$ & 96$^{+9}_{-9}$ & 170$^{+20}_{-36}$ & 0.64$^{+0.11}_{-0.12}$ & 1.18$^{+0.19}_{-0.21}$ \\
SPT\,0553--50 & 5.3201 & $s$           & 57.8$^{+7.1}_{-8.8}$  & 2.00$^{+0.21}_{-0.19}$ & 79$^{+5}_{-4}$ & 158$^{+44}_{-55}$ & 3.13$^{+0.26}_{-0.23}$ & 4.69$^{+0.46}_{-0.60}$ \\
HELMS45       & 5.3994 & $s$           & 92.9$^{+4.7}_{-4.7}$  & 2.19$^{+0.31}_{-0.32}$ & 55$^{+3}_{-2}$ & 187$^{+23}_{-22}$ & 7.19$^{+0.26}_{-0.25}$ & 16.3$^{+0.64}_{-1.81}$ \\
SPT\,2353--50 & 5.5781 & $s_{\rm c}$     & 72.0$^{+4.6}_{-5.5}$  & 2.53$^{+0.35}_{-0.32}$ & 71$^{+5}_{-4}$ & 220$^{+26}_{-34}$ & 3.28$^{+0.24}_{-0.21}$ & 5.62$^{+0.54}_{-0.70}$ \\
SPT\,0245--63 & 5.6256 &               & 55.7$^{+6.6}_{-18.5}$ & 2.38$^{+0.21}_{-0.19}$ & 88$^{+12}_{-7}$ & 179$^{+36}_{-64}$ & 3.78$^{+0.26}_{-0.25}$ & 18.2$^{+1.42}_{-1.26}$ \\
SPT\,0348--62 & 5.6541 & 1.18$\pm$0.01 & 64.4$^{+3.6}_{-5.2}$  & 2.65$^{+0.28}_{-0.21}$ & 79$^{+5}_{-4}$ & 197$^{+17}_{-29}$ & 3.36$^{+0.26}_{-0.21}$ & 5.35$^{+0.49}_{-0.64}$ \\
ADFS-27	      & 5.6550 &               & 59.2$^{+3.3}_{-4.1}$  & 2.52$^{+0.19}_{-0.17}$ & 85$^{+5}_{-4}$ & 191$^{+11}_{-19}$ & 1.58$^{+0.10}_{-0.19}$ & 2.38$^{+0.23}_{-0.22}$ \\
SPT\,0346--52 & 5.6554 & 5.6$\pm$0.1   & 75.9$^{+1.9}_{-2.2}$  & 2.37$^{+0.22}_{-0.23}$ & 67$^{+2}_{-2}$ & 195$^{+12}_{-14}$ & 12.01$^{+0.34}_{-0.27}$ & 20.41$^{+0.65}_{-0.67}$ \\
CRLE	      & 5.6666 & 1.09$\pm$0.02 & 62.6$^{+3.9}_{-3.5}$  & 1.61$^{+0.44}_{-0.45}$ & 76$^{+2}_{-2}$ & 180$^{+39}_{-44}$ & 1.61$^{+0.06}_{-0.06}$ & 2.62$^{+0.40}_{-0.25}$ \\
SPT\,0243--49 & 5.7022 & 5.1$\pm$0.5   & 41.9$^{+10.0}_{-10.8}$ & 2.16$^{+0.24}_{-0.24}$ & 101$^{+6}_{-6}$ & 155$^{+64}_{-73}$ & 3.66$^{+0.27}_{-0.27}$ & 7.59$^{+1.82}_{-2.03}$ \\
SPT\,2351--57 & 5.8114 & $s_{\rm c}$     & 85.6$^{+4.1}_{-5.0}$  & 2.90$^{+0.53}_{-0.57}$ & 60$^{+3}_{-3}$ & 232$^{+28}_{-39}$ & 3.92$^{+0.23}_{-0.18}$ & 7.74$^{+0.61}_{-0.70}$ \\
ID85001929    & 5.847  &               & 59.0$^{+7.7}_{-16.7}$ & 2.49$^{+0.04}_{-0.38}$ & 68$^{+5}_{-2}$ & 117$^{+11}_{-72}$ & 0.58$^{+0.08}_{-0.07}$ & 0.87$^{+0.10}_{-0.14}$ \\
HeLMS-54ab    & 5.880  &               &                    &                &         &           &                & \\
G09-83808     & 6.0269 & 8.2$\pm$0.3   & 57.4$^{+8.7}_{-8.9}$  & 2.65$^{+0.34}_{-0.33}$ & 83$^{+6}_{-6}$ & 174$^{+49}_{-47}$ & 2.44$^{+0.33}_{-0.32}$ & 3.61$^{+0.52}_{-0.58}$ \\
J1353$-$0010  & 6.1694 &               & 75.1$^{+21.6}_{-21.7}$ & 1.90$^{+0.52}_{-0.53}$ & 60$^{+11}_{-11}$ & 109$^{+43}_{-64}$ & 0.69$^{+0.13}_{-0.13}$ & 1.39$^{+0.55}_{-0.59}$ \\
HFLS3	      & 6.3369 & 1.8$\pm$0.6   & 63.3$^{+5.4}_{-5.8}$  & 1.94$^{+0.07}_{-0.09}$ & 73$^{+2}_{-1}$ & 142$^{+25}_{-27}$ & 2.93$^{+0.14}_{-0.13}$ & 5.50$^{+0.30}_{-0.22}$ \\
SPT\,0311--58 & 6.9011 & 2.0$\pm$0.2   & 80.4$^{+8.4}_{-8.3}$  & 2.51$^{+0.41}_{-0.40}$ & 64$^{+6}_{-6}$ & 213$^{+33}_{-35}$ & 3.97$^{+0.40}_{-0.37}$ & 8.32$^{+0.94}_{-1.71}$ \\
\tableline
median        & 5.36$\pm$0.27 &        & 63.9$\pm$9.0       & 2.52$\pm$0.21      & 73$\pm$8     & 189$\pm$31      & & \\
average       & 5.52$\pm$0.44 &        & 66.6$\pm$12.7      & 2.41$\pm$0.43      & 74$\pm$12    & 185$\pm$47      & & \\
\enddata
\tablenotetext{a}{Lensing magnification factor. Source thought to be strongly lensed but with unknown magnification factors (unlensed) are indicated with an ``s'' (empty column). An index ``c'' indicates that the source was identified as a cluster-lensed system by \citet{spilker16}.}
\tablenotetext{b}{Apparent values not corrected for gravitational magnification. $L_{\rm FIR}$ ($L_{\rm IR}$) is integrated over 42.5--122.5\,$\mu$m (8--1000\,$\mu$m) in the rest frame.}
\tablecomments{Photometry, redshifts and $\mu_{\rm L}$ used in the analysis are adopted from this work and \citet{reuter20} for all SPT sources, \citet{ikarashi22} for SXDF1100.053 (a.k.a.\ ASXDF1100.053.1), \citet{riechers21a} for ADFS-27, \citet{riechers21b} for HXMM-30, HeLMS-34, and HLock-102, \citet{cox23} for HELMS45, this work for J1353$-$0010, and \citet{riechers20a} for the remainder of the sample. The latter compilation includes additional data from \citep{rawle14,riechers10a,riechers13b,riechers14b,riechers17,pavesi18a,jin19,fudamoto17,zavala18}. Fluxes used for J1353$-$0010 at (3.4, 4.6, 12.1, 22.2, 635, 651, 1132, and 1195) $\mu$m are:\ ($<$0.0261, $<$0.0187, $<$0.0756, $<$5.28, 8.5$\pm$1.2, 8.09$\pm$0.42, 3.00$\pm$0.18, and 2.19$\pm$0.11) mJy, respectively, based on archival WISE and ALMA data. The near-infrared fluxes reported by \citep{mitsuhashi23} are only used as upper limit constraints to the Wien side for the dust SED. Other sources reported by these authors are too faint to make the luminosity cut of the sample considered here.}
${}$\\[-12.5mm]
\end{deluxetable*}

\subsection{SPT\,0452--50 as a $z$$>$5 DSFG}

\subsubsection{Physical Properties from Molecular Line Strengths}

As a proxy for the gas excitation, we can investigate the \eco\ to
\bco\ line brightness temperature ratio $r_{52}$ of
SPT\,0452--50. Given the impact of the warmer CMB at higher redshift,
direct comparisons only make sense to sources at similar
redshifts. For SPT\,0452--50, we find $r_{52}$=0.47$\pm$0.11. For the
three $z$=5.2--5.3 DSFGs HDF\,850.1, AzTEC-3, and GN10,
\citet{riechers20a} have found $r_{52}$=0.54$\pm$0.11, 0.78$\pm$0.07,
and 0.47$\pm$0.11, respectively. For the $z$=5.2 DSFG HLS\,0918, one
can find $r_{52}$=0.41$\pm$0.03 based on the fluxes reported by
\citet{rawle14}. For the $z$=5.7 DSFG ADFS-27, \citet{riechers21a}
have found $r_{52}$=0.60$\pm$0.05. For the $z$=5.7 DSFG CRLE,
\citet{vieira22} have found $r_{52}$=0.71$\pm$0.07. The median
(average) for this sample is $r_{52}$=0.54$\pm$0.07
($r_{52}$=0.57$\pm$0.14), where the uncertainties are the median
absolute deviation and the standard deviation, respectively. As such,
the $r_{52}$ of SPT\,0452--50 appears to be indicative of a rather
``typical'' CO excitation for a $z$$>$5 DSFG. This finding may call
the previously found strength (spectra shown by
\citealt{weiss13,reuter20}) of the line now identified as \fco\ into
question, but since no flux was reported for this line previously, we
cannot conclusively investigate this issue further.

\citet{yang16} have presented a relation between the
H$_2$O($J_{K_aK_c}$=2$_{02}$$\to$1$_{11}$) line luminosity and $L_{\rm
  IR}$, with a near-linear power law slope of
1.06$\pm$0.19. SPT\,0452--50 follows this relation, which is
consistent with a picture in which infrared pumping contributes
substantially to the excitation of the rotational H$_2$O lines (see
also, e.g., \citealt{ga12,riechers13b}). Its H$_2$O properties thus
appear typical for a massive DSFG.

From the \bco\ luminosity of SPT\,0452--50 at its revised redshift of
$z$=5.016, we find a lensing-corrected total molecular gas mass of
$M_{\rm gas}$($\alpha_{\rm CO}$)=1.3$\times$10$^{11}$\,\msol, where
$\alpha_{\rm CO}$=1.0\,\msol\,(K\,\kms\,pc$^2$)$^{-1}$ is the adopted
conversion factor from CO luminosity to gas mass (e.g.,
\citealt{riechers13b}). We here neglected the excitation correction
from \bco\ to \aco, as we consider it minor compared to other sources
of uncertainty. From $L_{\rm IR}$, we find a dust-obscured massive
star formation rate of SFR$_{\rm IR}$=4100\,\msol\,yr$^{-1}$. This
yields a gas depletion time of $M_{\rm gas}$/SFR$_{\rm IR}$=32\,Myr,
which is rather typical for a massive DSFG.

\subsubsection{[CII]/[OI] Luminosity Ratio}

The ratio between the fraction of [CII] to emerge from the neutral
interstellar medium and [OI] is an indicator of the density of
photon-dominated regions (PDRs; e.g., \citealt{luhman03}). We find a
luminosity ratio of $r_{\rm CII,OI}$=6.8$\pm$1.3 for
SPT\,0452--50. For the $z$=5.3 and 5.7 DSFGs AzTEC-3 and CRLE,
\citet{pavesi16,pavesi18a} find that 86\%$\pm$5\% and 84\%$\pm$4\% of
the [CII] emission come from PDRs based on their [CII]/[NII] line
ratios, respectively. Thus, we here assume a neutral fraction of 85\%
for the [CII] emission in SPT\,0452--50, which provides a ``neutral''
line luminosity ratio of $r_{\rm CIIn,OI}$=5.8$\pm$1.4, where we
increased the uncertainty by 20\% to account for the uncertainty in
the [CII] neutral fraction. Based on the dust optical depth of
SPT\,0452--50 at the respective line frequencies, we expect this ratio
to only be mildly affected by dust optical depth effects. For a
typical radiation field strength $G_0$ of order 10$^3$, the models by
Luhman et al.\ suggest a PDR density of order
$\sim$3--4$\times$10$^3$\,cm$^{-3}$ at this line ratio, which is not
unusual for star-forming gas.

\subsubsection{Is SPT\,0452--50 a ``Typical'' $z$$>$5 DSFG?}

\begin{figure*}
\epsscale{1.0}
\plottwo{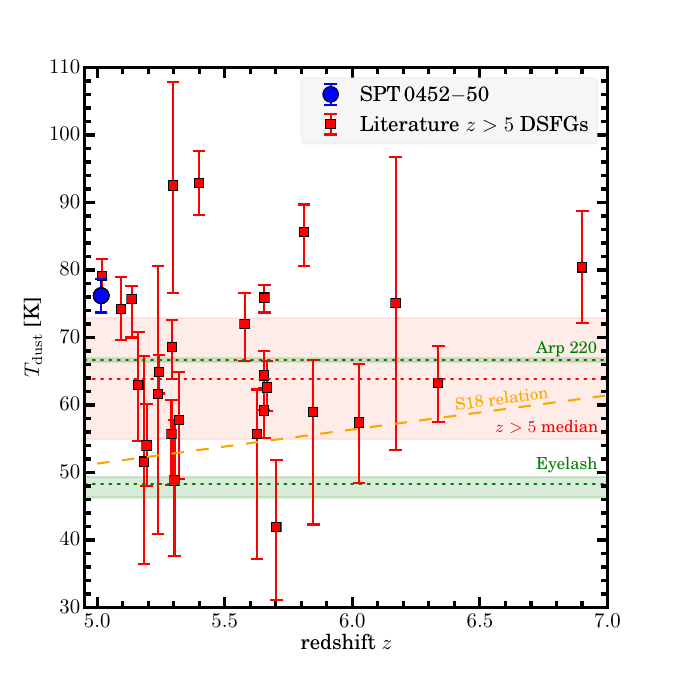}{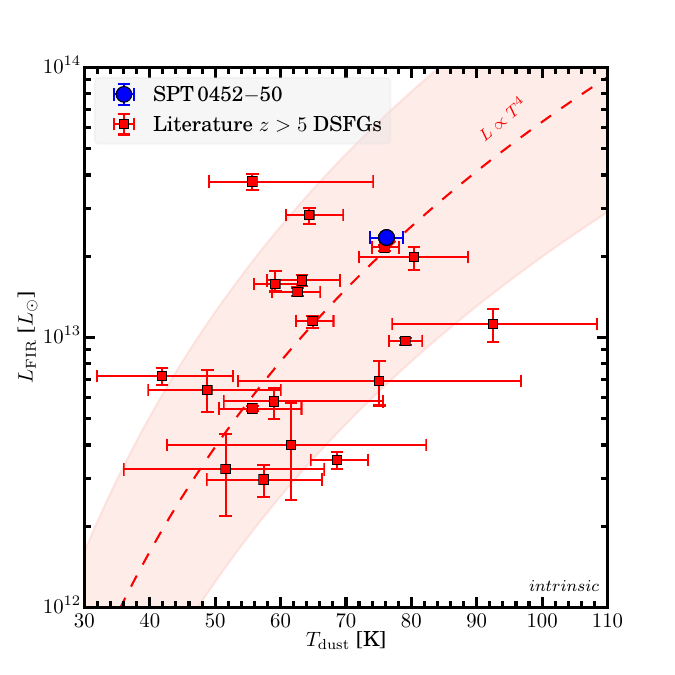}
\vspace{-2mm}

\caption{$T_{\rm dust}$ -- redshift relation (left) and $T_{\rm dust}$
  -- $L_{\rm FIR}$ relation (right), updated from
  \citet{riechers20a}. Only sources with reliable magnification
  factors are included in the right panel, and corrected
  correspondingly. Left:\ For reference, the median redshift of the
  sample and the values for the dusty starbursts Arp\,220 and the
  Eyelash are shown as dotted lines, where the shaded regions indicate
  the median absolute deviation and 1$\sigma$ uncertainties,
  respectively. In addition, the scaling relation proposed by
  \citet{schreiber18} is shown as a dashed line. Right:\ A standard $L
  \propto T^4$ scaling, normalized to the sample median, is shown for
  reference as a dashed line, with a $\pm$0.5\,dex range added as a
  shaded region. \label{f7}}
%
\end{figure*}

Since SPT\,0452--50 was an outlier in the relationships between
$T_{\rm dust}$, $L_{\rm FIR}$ and $z$ in the SPT sample as reported by
\citet{reuter20}, we here return to the question if it remains unusual
after revising the redshift. In the left panel of Figure~\ref{f7}, the
$T_{\rm dust}$--$z$ relation for all known $z$$>$5 DSFGs is shown. The
difference in values for some sources compared to the compilation by
\citet{riechers20a} is due to the updated SED fits for those sources.

The relation shows that the median $T_{\rm dust}$ of known $z$$>$5
DSFGs is more similar to Arp\,220 (and thereby, the $z$=6.34 DSFG
HFLS3) than to the Cosmic Eyelash when fit with the same SED code,
although 10\%--15\% of the sample have a $T_{\rm dust}$ close to the
Eyelash. This is interesting, because the above sources are commonly
used as SED templates for finding the most distant DSFGs. While there
is no obvious trend of $T_{\rm dust}$ with $z$ within the sample, we
also find that the revised median $T_{\rm dust}$ appears consistent
with the $T_{\rm dust}$--$z$ trend proposed by \citet{schreiber18},
but we caution that the SED fitting methods differ.

In the right panel of Figure~\ref{f7}, the relationship between
$T_{\rm dust}$ and $L_{\rm FIR}$ is shown for the same sample. This
panel contains fewer points, because sources with unknown lensing
magnification factors were removed. With the updated SED fits compared
to previous work, the trend is even more consistent with a standard $L
\propto T^4$ scaling relation.

In combination, these relations show that SPT\,0452--50 is a rather
``typical'' $z$$>$5 DSFG among the known specimens, where its
higher-than-average $T_{\rm dust}$ can be entirely explained by its
relatively high intrinsic $L_{\rm FIR}$. As such, it is no longer an
outlier, and most certainly not an unusually cold starburst.

\subsubsection{A $z$$>$5 DSFG SED Template}

The median (average) redshift of the full $z$$>$5 DSFG sample is
5.36$\pm$0.27 (5.52$\pm$0.44), where the uncertainties are the median
absolute deviation and the standard deviation, respectively. Based on
our analysis, we provide a median SED template for $z$$>$5 DSFGs, with
the parameters $T_{\rm dust}$=63.9$\pm$9.0\,K, $\beta_{\rm
  IR}$=2.52$\pm$0.21, $\lambda_{\rm peak}$=73$\pm$8\,$\mu$m, and
$\lambda_0$=189$\pm$31\,$\mu$m in the rest frame. A corresponding
average-based SED template would take the form $T_{\rm
  dust}$=66.6$\pm$12.7\,K, $\beta_{\rm IR}$=2.41$\pm$0.43,
$\lambda_{\rm peak}$=74$\pm$12\,$\mu$m, and
$\lambda_0$=185$\pm$47\,$\mu$m, which is indistinguishable within the
uncertainties. The corresponding median (lensing-corrected) $L_{\rm
  FIR}$ of the template is (1.05$\pm$0.56)$\times$10$^{13}$\,\lsol,
with an average of (1.27$\pm$0.95)$\times$10$^{13}$\,\lsol.

A potential fine-tuning of the template could be to add a scaling of
$T_{\rm dust}$ with $L_{\rm FIR}$, of the form $L_{\rm
  FIR}$=$a$$T_{\rm dust}^4$. Scaled to the median, we find
$a$=5.80. We caution that this correction will, of course, not be
valid for samples where strong lensing is expected to be a major
contributor to the selection.

\subsection{New Constraints on the Gas Properties of SPT-Selected DSFGs}

\subsubsection{\ico--$L_{\rm FIR}$ Relation}

\begin{figure*}[tbh!]
\epsscale{1.0}
\plottwo{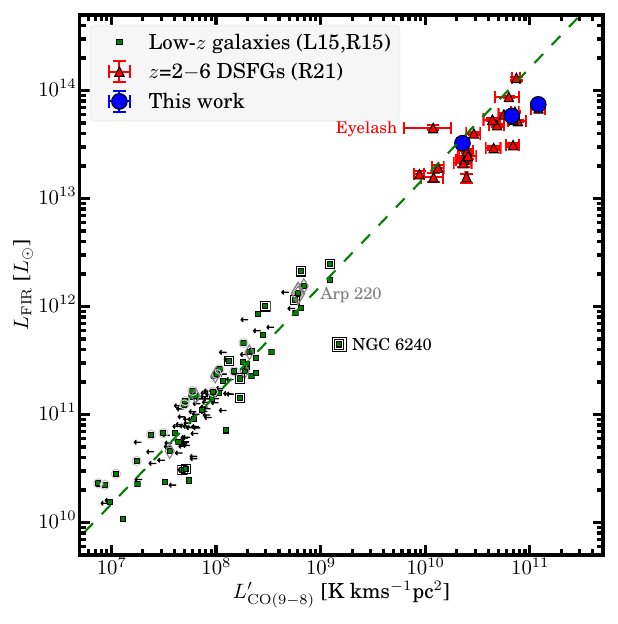}{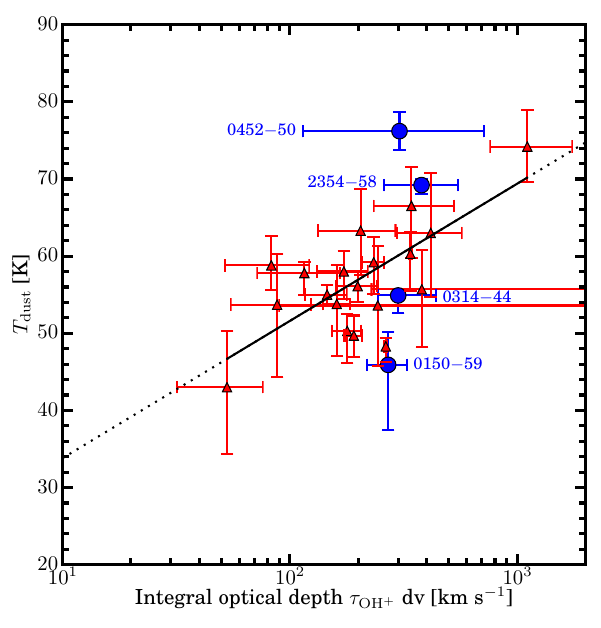}
\vspace{-2mm}

\caption{Left:\ \ico -- FIR luminosity relation, updated from
  \citet{riechers21b}, including nearby galaxies (green symbols and
  upper limit arrows; \citealt{liu15}; L15), a high--$z$ DSFG sample
  from the literature (red triangles; all data from
  \citealt{riechers21b}; R21), and the sources reported here (blue
  dots). The dashed line is the power-law fit to the nearby galaxy
  sample from L15, with a slope of $N$ = 1.01. Luminosities were not
  corrected for gravitational lensing magnification. Gray circle,
  diamond, and square outlines indicate Class I, II, and III sources,
  respectively, where higher classes indicate higher CO excitation
  (\citealt{rosenberg15}; R15). Right:\ Relation (black line) between
  the dust temperature and the OH$^+$ integral optical depth, with the
  same marker styles, updated from \citet{riechers21b}. The relation
  has the functional form $T_{\rm dust}=a+1\,{\rm K}\,{\rm
    log}(b\,\tau_{\rm OH+}{\rm dv})$, where $a$=(16.0$\pm$12.5)\,K and
  $b$=(7.7$\pm$2.3)(\kms)$^{-1}$. \label{f8}}
%
\end{figure*}

\citet{riechers21b} have found a strong correlation between the
\ico\ and FIR dust continuum luminosities for a sample of 20 highly
luminous, {\em Herschel}-selected starburst galaxies at $z$=2--6 and
the Cosmic Eyelash at $z$=2.3, consistent with the idea that the
\ico\ emission is associated with warm and dense molecular gas in the
star-forming regions. The three SPT-selected galaxies with detections
reported here fall within the scatter of the {\em Herschel} sample,
and thus, appear indistinguishable in their properties for this
relation (Fig.~\ref{f8}, left). As such, the combined sample appears
systematically offset toward higher \ico\ luminosity when compared to
nearby star-forming galaxies, strengthening the trend reported by
\citet{riechers21b}. As explained by these authors, this is likely
either due to increased shock excitation, increased cosmic-ray energy
densities, or a combination of both effects.

\subsubsection{$T_{\rm dust}$--OH$^+$ opacity relation}

\citet{riechers21b} have also reported an apparent relation between
the dust temperature and OH$^+$ integrated optical depth $\tau_{\rm
  OH+}$d$v$ for the same sample of {\em Herschel}-selected
sources. This relation is likely related to an underlying relation
between $T_{\rm dust}$ and the star formation rate surface density,
$\Sigma_{\rm SFR}$, which yields an increased cosmic ray energy
density in more compact, warmer sources due to a higher supernova
density. Since higher cosmic ray fluxes increase the OH$^+$ ion
abundance, this then can lead to an increase in OH$^+$ absorption
strength.

For the two new SPT sources where both OH$^+$ 1$_1$$\to$0$_1$ and
1$_2$$\to$0$_1$ have been detected, we find that the $\tau_{\rm
  OH+}$d$v$ between both lines are indistinguishable within the
uncertainties, consistent with what has been reported by
\citet{riechers21b}. As such, we also include SPT\,0452--50 in this
analysis, where only the 1$_2$$\to$0$_1$ transition has been
measured. Three of the SPT galaxies lie within the scatter of the
previous data points, with the lowest signal-to-noise detection
SPT\,0452--50 showing the largest distance from the trend. The modest
increase in sample size neither substantially strengthens or weakens
the previous trend (Fig.~\ref{f8}, right).

The average column density of $N({\rm OH^+})$ of
(1.5$\pm$0.2)$\times$10$^{15}$\,cm$^2$ for the new sources is
consistent with the median reported for the larger sample studied by
\citep{riechers21b}. Combining both samples, we find a revised median
value of (1.1$\pm$0.4)$\times$10$^{15}$\,cm$^2$, which corresponds to
a median atomic hydrogen column density of
(7.2$\pm$2.2)$\times$10$^{22}$\,cm$^2$.

\subsubsection{Comparing NH and OH$^+$ Absorption}

NH($N_J$=1$_2$$\to$0$_1$) was previously detected in absorption in the
$z$=2.95 DSFG HerBS-89a, and in emission in the $z$=3.39 DSFG Orochi
(\citealt{berta21,riechers21b}). The absorption lines detected here
toward SPT\,0150--59 and 0314--44 show $\sim$20\%--40\% of the
integrated optical depth of OH$^+$($N_J$=1$_2$$\to$0$_1$), which is
similar to the $\sim$20\% seen in HerBS-89a. Given the modest optical
depths seen for both species, the difference in line strengths between
NH and OH$^+$ is likely a good indicator of the difference in
abundances of both molecules in the same environments.

\section{Conclusions} \label{sec:conclusions}

We have targeted the \ico\ and OH$^+$($N_J$=1$_1$$\to$0$_1$) line in
the four millimeter-selected, high-redshift dusty star-forming
galaxies SPT\,2354--58, 0150--59, 0314--44, and 0452--50. For the
first three sources, we independently confirm their redshifts and find
line and dust continuum properties consistent with the relations
between \ico\ luminosity and far-infrared luminosity and between the
OH$^+$ integral optical depth and dust temperature proposed by
\citet{riechers21b}. These findings are consistent with the presence
of dust-enshrouded starbursts permeated by a high cosmic ray energy
density leading to high \ico\ luminosities and OH$^+$ abundances, with
a possible contribution by increased shock excitation to the CO line
ladder.

Based on a serendipitous \oi\ detection in these data, and in
combination with follow-up, archival, and literature data, we also
find that the last source, SPT\,0452--50, is not an unusually cold
massive starburst at $z$=2.011 as previously thought, but rather a
hyper-luminous massive $z$=5.016 starburst with properties typical for
this population.

We analyze SPT\,0452--50 in concert with the known $z$$>$5 dusty
starburst population, and propose a simple dust spectral energy
distribution template which may be useful for the identification of
more such systems at $z$$>$5, and for the extraction of physical
properties for similar systems with limited photometry.

Our analysis does not provide evidence for the existence of unusually
cold starburst galaxies in the early universe which could have been
missed by some of the standard selection techniques. While they may
still exist, it remains unclear that they are a major source of bias
for the study of $T_{\rm dust}$--$z$ relations for dust-selected
samples.

\begin{acknowledgments}

D.R. gratefully acknowledges support from the Collaborative Research
Center 1601 (SFB 1601 sub-projects C1, C2, C3, and C6) funded by the
Deutsche Forschungsgemeinschaft (DFG) – 500700252. This work makes use
of the following ALMA data: ADS/JAO.ALMA\#\,2023.1.01481.S,
2021.2.00062.S, 2019.1.00297.S, 2019.1.01026.S, 2016.1.00231.S,
2011.0.00957.S, and 2011.0.00958.S. ALMA is a partnership of ESO
(representing its member states), NSF (USA) and NINS (Japan), together
with NRC (Canada) and NSC and ASIAA (Taiwan), in cooperation with the
Republic of Chile.

\end{acknowledgments}

\vspace{5mm}
\facilities{ALMA}

\bibliography{ms}{}
\bibliographystyle{aasjournal}


\end{document}